\documentclass[12pt]{JHEP3}
\newcommand{\Yb}{\ensuremath{\overline{Y}}}
\newcommand{\yh}{\ensuremath{\hat{y}}} 
\newcommand{\htw}{\ensuremath{\tilde{h}}} 
\newcommand{\Ah}{\ensuremath{\hat{A}}}

\newcommand{\g}{\ensuremath{\gamma}}
\newcommand{\eref}[1]{(\ref{#1})}
\def\be{\begin{equation}}
\def\ee{\end{equation}}
\def\ba{\begin{eqnarray}}
\def\ea{\end{eqnarray}}

\title{Intersecting $D$--branes, Chern--kernels and the inflow mechanism}  

\author{Marco Cariglia  
        \\ 
        DAMTP, Centre for Mathematical Sciences, Cambridge University
        \\ Wilberforce Road, Cambridge CB3 OWA, UK \\ 
         E-mail: \email{m.cariglia@damtp.cam.ac.uk}} 
\author{Kurt Lechner  
        \\ 
        Dipartimento di Fisica, Universit\`a degli Studi di Padova, \\
	and \\ 
	Istituto Nazionale di Fisica Nucleare, Sezione di Padova, via
        F. Marzolo 8, 35131 Padova, Italia \\ 
        E-mail: \email{kurt.lechner@pd.infn.it}}

\preprint{DAMTP-2004-44 \\ DFPD 04/TH/03}

\abstract{We analyse a system of arbitrarily intersecting 
$D$--branes in ten--dimensional supergravity. Chiral anomalies are supported on the intersection branes, called 
$I$--branes. For non--transversal intersections 
anomaly cancellation has been realized until now
only cohomologically but not locally, due to  
short--distance singularities. In this paper we present a consistent local cancellation 
mechanism, writing the $\delta$--like brane currents as 
differentials of the recently introduced Chern--kernels, $J=dK$.  
In particular, for the first time we achieve anomaly cancellation 
for dual pairs of $D$--branes. The Chern--kernel approach allows to construct an effective 
action for the $RR$--fields which is 
free from singularities and cancels the quantum anomalies on all 
$D$--branes and $I$--branes.}

 
\begin{document}
 \section{Introduction and summary}

Anomaly cancellation represents a basic quantum--consistency check 
for extended objects in $M$--theory, 
and constrains eventually the physically allowed excitations. 
A particularly interesting case regards $D$--branes in ten dimensions.
$IIA$--branes carry an odd--dimensional worldvolume and are trivially 
anomaly--free, while $IIB$--branes carry an even--dimensional worldvolume
and are actually plagued by gravitational anomalies; the problem of their
cancellation has been addressed for the first time in \cite{cinesi}.

However, $D$--branes may also intersect with each other. The requirement 
of anomaly freedom for such configurations has indeed been used in \cite{Green}
to deduce the anomalous (Wess--Zumino) couplings of the $IIA$ and $IIB$ 
$RR$--potentials to the fields on the branes: if the intersection manifold, 
called $I$--brane, is even--dimensional there are potential anomalies supported
on it, which have to be cancelled by adding specific Wess--Zumino terms
to the action and by modifying, correspondingly, the Bianchi--identities 
for $RR$--curvatures. Anomalies on $I$--branes represent
the main topic of the present paper.

One has to distinguish two kinds of $I$--branes.
In a $D$--dimensional spacetime two $D$--branes with $i$--dimensional 
worldvolume $M_i$ and, respectively, $j$--dimensional
worldvolume $M_j$ may indeed intersect in two
different ways, depending on the dimension of the $I$--brane manifold 
$M_{ij}=M_i\cap M_j$. In the first (generic) case we have
$$
dim(M_{ij})=i+j-D,
$$ 
and the intersection is called {\it transversal}; for
such intersections the normal bundles of the two branes do not intersect,
$N_{ij}\equiv N_i\cap N_j=\emptyset$. If $i+j-D<0$ it is understood that
$M_{ij}=\emptyset$. An example of a transversal intersection are two planes 
in three dimensions that intersect along a line. 

In the second (exceptional) case the dimension of the $I$--brane satisfies 
$$ 
dim(M_{ij})>i+j-D,
$$
and the intersection is called {\it non--transversal}. In this case
$N_{ij}\neq\emptyset$ and $dim(N_{ij})= dim(M_{ij})+D-(i+j)$.
If $i+j-D<0$ it is understood that $M_{ij}\neq\emptyset$.
Examples in three dimensions are two coinciding
planes, or two lines which intersect in a point. 

The anomaly cancellation mechanism presented in \cite{Green} applies
to transversal $I$--branes, while the attempt of \cite{cinesi} was to
include the case of non--transversal $I$--branes as well. Actually, 
the anomaly cancellation mechanism of \cite{cinesi} achieves only a
``cohomological'' cancellation, in a sense that we will specify more 
precisely in a moment.
The main purpose of the present paper is to fill this gap, i.e. to present 
a cancellation mechanism for non--transversal intersections which works 
``locally'', point--wise, as explained below.

The difference between these two kinds of $I$--branes can be translated 
in the language of differential forms as follows.
Introduce the $\delta$--function supported Poincar\`e dual 
forms for $M_i$ and $M_j$, i.e. their ``currents'' $J_i$ and $J_j$, of degree 
$D-i$ and $D-j$ respectively, and the current $J_{ij}$ associated to 
$M_{ij}$, of degree $D-dim(M_{ij})$. Then for transversal intersections the 
product $J_iJ_j$ is well--defined and one has simply 
\be
J_iJ_j=J_{ij}.
\label{1}
\ee
For non--transversal intersections the degree of the product $J_iJ_j$ 
is greater then the degree of  $J_{ij}$ but, moreover, the product
itself is ill--defined.
The reason is that since $N_{ij}$ is non empty, there
exists at least one direction in $N_{ij}$, parametrized by a coordinate 
say $u$, such that $J_i$ as well as $J_j$ contain a factor $du\,\delta(u)$.
The product $J_iJ_j$ is therefore of the kind 0 $(du\wedge
du)$ times $\infty$ $(\delta(u)\delta(0))$.

For what concerns anomalies on $I$--branes, the obstacle to their 
cancellation on non--transversal intersections arises as follows.
Suppose first that the intersection is transveral. Then the anomaly 
polynomial due to 
chiral fermions on the $I$--brane $M_{ij}$ is nonvanishing, and it amounts to 
a sum of factorized terms \cite{cinesi,Green},
\be
P_{ij}=2\pi P_iP_j,   
\ee
$P_i(P_j)$ being supported on $M_i(M_j)$.  The anomaly is given by the 
descent \footnote{As usual for a generic polynomial we set 
$P=dP^{(0)}$, $\delta P^{(0)}=dP^{(1)}$.}
\be
\label{2}
{\cal A}=2\pi \int _{M_{ij}}(P_iP_j)^{(1)}.
\ee
It is cancelled by a Wess--Zumino term in the action  of the form
\be
\label{2a}
S_{WZ}=2\pi\int_{M_i} P_i \widetilde{C}     ,
\ee
where the $RR$--potential $\widetilde{C}$ entails an anomalous transformation
supported on $M_j$, 
\be
\label{tra}
\delta \widetilde{C}=-P_j^{(1)}J_j.
\ee 
The $WZ$--term varies according to
\be
\delta S_{WZ}=-2\pi\int_{M_i} P_j^{(1)}J_jP_i=-2\pi\int J_iJ_j(P_iP_j)^{(1)}, 
\label{3}
\ee
which cancels ${\cal A}$ thanks to \eref{1}.

For non--transversal intersections the addenda in the anomaly polynomial 
factorize only partially \cite{cinesi}
$$
P_{ij}=2\pi P_iP_j\,\chi_{ij},
$$
due to the presence of the Euler--form $\chi_{ij}\equiv \chi(N_{ij})$ 
of the now no longer 
vanishing intersection of the normal bundles $N_{ij}$ -- a form of degree
$dim(M_{ij})+D-(i+j)$. The anomaly reads then 
\be
\label{4}
{\cal A}= 2\pi\int _{M_{ij}}\chi_{ij}(P_iP_j)^{(1)}.
\ee
The Wess--Zumino term is still given by \eref{2a} but its
{\it formal} variation leads to the now ill--defined expression \eref{3}.

We can now make a more precise statement of the problem attacked 
in this paper and outline its solution. For non--transversal intersections
the quantum anomaly \eref{4} is still well--defined, what is ill--defined
is the Wess--Zumino term \eref{2a} itself, since its variation leads
to an ill--defined expression. 
The problem consists therefore in constructing a 
well--defined Wess--Zumino term and, afterwards, in checking whether its
variation cancels \eref{4} or not.

The authors of \cite{cinesi} proposed a partial solution to this problem,
maintaining the above Wess--Zumino term together with its variation \eref{3},
and trying to give a meaning to 
the product ``$J_iJ_j$''. Clearly, to cancel the anomaly what one would need is
the identification
\be
\label{4a}
J_iJ_j\leftrightarrow J_{ij}\chi_{ij}.
\ee
As it stands this identification is rather contradictory since
$J_iJ_j$ is simply a product of $\delta$--functions, while the r.h.s. contains,
apart from $\delta$--functions, the gravitational curvatures present in
$\chi_{ij}$. Moreover, the l.h.s. is ill--defined.  
The authors of \cite{cinesi} proposed first to substitute
say $J_i$ by a smooth {\it cohomological} representative $\hat J_i$. Then 
they showed that the product $\hat J_i J_j$ is {\it cohomologically 
equivalent to} $J_{ij}\chi_{ij}$, in the sense of de Rham. 
Although this is clearly not enough to realize
a local anomaly cancellation mechanism, the above identification bears 
convincingly the correct idea.  
 
The main lines of our solution are, indeed, as follows. The expression 
\eref{2a} itself looks canonical and rigid: the unique feature one can 
try to change is 
the definition of the $RR$--potential $\widetilde{C}$. Above it is indeed (implicitly)
assumed that the $RR$--field strength is given in terms of $\widetilde{C}$ as
\footnote{We focus here only on the above anomaly; the complete Bianchi 
identity for a $RR$--curvature is more complicated, especially for the
presence of the $B_2$-field, and it is given in the text.}
\be
 dR=P_jJ_j \leftrightarrow  R=d\widetilde{C}+P_j^{(0)}J_j,
\label{5}
\ee
which obliges $\widetilde{C}$ to the transformation law \eref{tra}, carrying
a $\delta$--like
singularity on $M_j$, meaning that $\widetilde{C}$ itself is singular
on $M_j$, and therefore that \eref{2a} is ill-defined. Our strategy instead consists in keeping \eref{2a}, 
while introducing a $RR$--potential that is regular on $M_j$, actually on all 
branes. 
A key step in this direction is to search for a convenient antiderivative 
$K_j$ of the current,
\be
\label{5a}
J_j=dK_j.
\ee 
Then one can solve the Bianchi identity for $R$ alternatively in terms of 
a different potential
\be
\label{5b}
R=d C +P_jK_j, \quad \widetilde{C}= C - P_j^{(0)}K_j,
\ee
where $C$ is invariant and, for a convenient choice of $K_j$,
regular on $M_j$. This time $\delta \widetilde{C}=-dP_j^{(1)}K_j$ and  
the variation of the Wess--Zumino term \eref{2a}
amounts to
$$ 
\delta S_{WZ}=-2\pi\int_{M_i}d\left(P_iP_j\right)^{(1)}K_j=
-2\pi\int d\left(J_iK_j\right)(P_iP_j)^{(1)}.
$$ 
The difference w.r.t. \eref{3} is that for a convenient choice of $K_j$ 
{\it the product $J_iK_j$ may be well--defined together with its differential}
-- contrary to what happens to $J_iJ_j$. The apparent paradox is solved
by the fact that, due to the singularities present, one is not allowed to use 
Leibnitz's rule when computing $d(J_iK_j)$.
 
The key observation of the present paper is that if one chooses for $K_j$ a 
Chern--kernel \cite{Harvey_Lawson} then the product is 
not only well--defined, but one also has the new fundamental identity
\be
\label{6}
d(J_iK_j)=J_{ij}\chi_{ij},
\ee
realizing in some sense the ``identification'' \eref{4a}, which is 
precisely what is needed to cancel the anomaly. 
To be precise, this formula holds whenever $M_i\not\subset M_j$.
The extremal case $M_i\subset M_j$ needs a slight adaptation that is given
in the text. For previous 
applications of the Chern--kernel--approach to anomaly cancellation
see \cite{Lechner_Marchetti}--\cite{Lechner}. 

A special case of non--transversal $I$--branes is represented by
a couple of electromagnetically dual branes,
$$
i+j=D-2,
$$
e.g. a $D1$-- and a $D5$--brane, which have a non--empty intersection 
manifold $M_{ij}$. Then the intersection of the normal bundles has dimension 
$dim(N_{ij})= dim(M_{ij})+2$, and if it is even it has a nonvanishing  
Euler--form of the same degree. In this case the anomaly polynomial 
on $M_{ij}$ is given simply by the Euler--form itself \cite{cinesi},
$$
P_{ij}=2\pi\chi_{ij},
$$ 
but its cancellation has not yet been achieved, not even 
``cohomologically''; for the solution of a particular example in
$D=11$ see however \cite{Lechner}.
As stated in \cite{cinesi},  the cancellation of these anomalies requires
``a more powerful approach'': Chern--kernels provide actually
such an approach. Indeed, in this case the relevant 
contribution in the Bianchi--identity, realizing the minimal coupling but
ignored in \cite{cinesi}, is 
\be
\label{66}
dR=J_j \leftrightarrow R=d C+K_j,
\ee
and the Wess--Zumino term is conveniently written as the integral
over an eleven--dimensional manifold, with space--time $R^{10}$ 
as boundary, of a closed eleven--form:
\be
\label{6a}
S_{WZ}= 2\pi\int_{M_{11}}\left(RJ_i-J_{ij}\chi_{ij}^{(0)}\right).
\ee
The eleven--form is closed thanks to \eref{6}, and 
$$
\delta S_{WZ}=-2\pi\int_{M_{11}} d\left(J_{ij}\chi_{ij}^{(1)}\right)
            =-2\pi\int_{R^{10}}J_{ij}\chi_{ij}^{(1)}
            =-2\pi\int _{M_{ij}}\chi_{ij}^{(1)},
$$
which cancels the anomaly. Again, as we will see the definition \eref{66} 
leads to a potential $C$ that is regular on $M_j$.

A third case regards the anomalies on the (even-dimensional) $D$--branes of 
$IIB$--supergravity. Formally the anomalies supported on a
$D$--brane can be interpreted as anomalies on the $I$--brane of two
copies of the same $D$--brane (self--intersection). In light of this 
interpretation these anomalies are just a special  case of anomalies on
$I$--branes (their cancellation has been discussed in  
\cite{cinesi}, again from a cohomological point of view). So the Chern--kernel
approach furnishes automatically a consistent local cancellation mechanism
also for $IIB$ $D$--branes.

For concreteness in this paper we consider a system of 
arbitrarily interacting and intersecting abelian $IIB$--branes (one for
each woldvolume dimension), 
the case of abelian $IIA$--branes requiring only a straightforward
adaptation. Actually, $IIB$--branes have a richer anomaly structure because,
being even dimensional, they carry anomalies even in the absence of
intersections. The generalization to non--abelian branes is exposed briefly
in the conclusions. 

Usually a magnetic equation of the kind $dR=J_j$ requires the introduction
of a Dirac--brane, as antiderivative of $J_j$, whose unobservability is 
guaranteed by charge quantization. The consistency of the employment 
of Chern--kernels as antiderivatives, instead of Dirac--branes, 
has been proven in \cite{Lechner_Marchetti}. 

In section two we recall the  
definition of odd ($IIB$) and even ($IIA$) Chern--kernels, and review in a 
self--contained way their basic properties. In section three we present 
the basis for the fundamental identity \eref{6}. In section four
we show how one arrives at formula \eref{6a} for a pair of dual branes,
explaining the interplay between Dirac--branes and  Chern--kernels.
In section five we recall the specific form of the  
anomalies produced by chiral fermions on $D$--branes and $I$--branes, 
we give the complete set of  Bianchi--identities/equations of motion 
for the $RR$--field strengths of $IIB$--Sugra in presence of branes, and
present their solutions in terms of Chern--kernels and regular potentials.
This section is based on a systematic application and elaboration of our 
proposals for the introduction of regular potentials, made in \eref{5b}
and \eref{66}.
We take also a non--vanishing $NS$ $B_2$--field into account, whose 
consistent inclusion is not completely trivial.
In this section we write eventually the action, in particular the 
Wess--Zumino term, producing the correct equations of motion (for the 
``basic'' potentials $C_0$, $C_2$ and 
$C_4$), verifying that it is well--defined and that it 
cancels all anomalies. Section six is more technical, in that there we 
write a manifestly duality--invariant (physically equivalent) action, in which 
the $RR$--potentials $C_0$, $C_2$, $C_4$ and their duals $C_6$, $C_8$ 
appear on the same footing. 
In this form the distinctive features of our action with respect to  
previous results emerge more clearly. Section seven is devoted to 
generalizations and conclusions.
 
We remark briefly on our conventions and framework. We
will assume that there are no topological obstructions in spacetime, in
particular closed forms in the bulk are then always 
exact. Since we are in presence of $\delta$--like currents,  
for consistency differential forms are intended as distribution--valued,
and the differential calculus is performed in the sense of
distributions. This implies that our differential operator $d$ is 
{\it always} nihilpotent, $d^2=0$. With our conventions it acts 
from the right rather than from the left.

\section{Chern--kernels: definition and properties \label{sec:Chern_kernels}} 

In this section we review briefly the  definition of Chern--kernels and recall
their main properties \cite{Lechner_Marchetti}. Since we will treat in 
detail only $IIB$--branes, that have an even--dimensional worldvolume,
we concentrate mainly on odd Chern--kernels, but for completeness and 
comparison we report also shortly on even kernels.
For more details we refer the reader to the above reference. 

\subsection{Odd kernels}

Let $M$ be a closed $(D-n)$--dimensional brane worldvolume in a
$D$--dimensional
spacetime, and introduce a set of normal coordinates $y^a$, $(a=1,\cdots,n)$
associated to $M$; the brane stays at $y^a=0$. Then locally one can write 
the current associated to $M$ as
\be
J = \frac{1}{n!} \, \varepsilon^{a_1\dots a_n} \, dy^{a_1} \dots
dy^{a_n} \, \delta^n(y).
\ee
One can also introduce an $SO(n)$--connection $A^{ab}$ and its
curvature $F=dA+AA$ (both are target--space forms), which are only 
constrained to reduce, if restricted to $M$, respectively to the 
$SO(n)$--normal--bundle connection and curvature, defined intrinsically 
on $M$. 

For odd rank Chern--kernels (even currents, $IIB$--branes)  
$n$ is even, $n=2m$. Then one can define the Euler $n$--form 
\footnote{In the following odd dimensional Euler forms are taken to be 
zero {\it by definition.}} associated to $F$ and its Chern--Simons form, 
$$
	\chi= \frac{1}{m!(4\pi)^m} \, \varepsilon^{a_1\dots a_n}
	\, F^{a_1 a_2}\dots F^{a_{n-1} a_n} = d \,\chi^{(0)}.
$$ 
Its anomaly descent is indicated as usal by 
$\delta \chi^{(0)}=d\chi^{(1)}$. Notice that the rank of the Euler--form
equals the rank of the current. The Chern--kernel $K$ associated to 
the even current $J$ is written as the sum
\ba 
	K&=& \Omega + \chi^{(0)}, \label{eq:K_def}
	\\ 
	dK&=&J,
\ea
where $\Omega$ is an {\it $SO(n)$--invariant $(n-1)$--form with 
inverse--power--like singularities on $M$}, polynomial in 
$F$ and $D\yh=d\yh-A\yh$, where $\yh^a=y^a/|y|$.
For the expression of $\Omega$ for a generic $n$ see 
\cite{Lechner_Marchetti}; for example for $n=2$ and $n=4$ respectively, 
the formula reported there gives 
\ba
\Omega &=&  -{1\over 2 \pi}\,\varepsilon^{a_1a_2}\yh^{a_1}D\yh^{a_2},\\
\Omega &=& -{1\over 2(4\pi)^2}\,\varepsilon^{a_1\dots
a_4}\yh^{a_1}D\yh^{a_2} \left( 4F^{a_3a_4}+
\frac{8}{3}D\yh^{a_3}D\yh^{a_4}\right).
\ea

Chern--kernels are not unique due to the arbitrariness of normal 
coordinates and of $A$ {\it away from the brane}, i.e. in the bulk, and due
to the presence of the non--invariant Chern--Simons form $\chi^{(0)}$. But 
since for a different kernel one has in any case $dK^\prime=J$, one obtains 
\be
\label{eq:Q_transform1} 
K^\prime = K+dQ,
\ee
for some target--space form $Q$. What matters eventually is the behaviour
of $Q$ on $M$. Since $\Omega$ has a singular but {\it invariant} behaviour
near the brane, it is only $\chi^{(0)}$ that induces an {\it anomalous 
but finite} change on $M$,
\be
\label{eq:Q_transform2}
Q|_M=\chi^{(1)}|_M.
\ee
The transformation \eref{eq:Q_transform1} has been called $Q$---transformation
in \cite{Lechner_Marchetti} and it is in some sense the analogous of a change
of Dirac--brane. From \eref{eq:Q_transform2} one sees that on $M$ a 
$Q$--transformation reduces to a normal--bundle
$SO(n)$--transformation, and it can give rise to anomalies supported on $M$.
On the contrary, we demand a theory to be {\it $Q$--invariant in the bulk}. 
As we will see below, also the $RR$--potentials have to transform under 
$Q$--transformations, and so $Q$--invariance and anomalies
are intimately related.  

\subsection{Even kernels}

For odd currents ($IIA$--branes) $n$ is odd, and the kernel is even. 
In this case it is only constructed from an {\it $SO(n)$--invariant $(n-1)$--form 
with inverse--power--like singularities on $M$}, ($K=\Omega$) 
\ba 
	K&=& {\Gamma\left(n/2\right) \over 2
\pi^{n/2}(n-1)!}\,\varepsilon^{a_1\cdots a_n}\,\yh^{a_1}\,{\cal
F}^{a_2a_3}\cdots {\cal F}^{a_{n-1}a_n}, \\ 
	dK&=&J,
\ea
with ${\cal F}^{ab} \equiv F^{ab}+D\yh^aD\yh^b$. The reason is that for an 
odd normal bundle the Euler form is vanishing. 

Also this kernel is defined modulo $Q$--transformations, 
$K^\prime=K+dQ$, but since $\Omega$ is invariant this time we have
$$
Q|_M=0.
$$

We can thus write in general $K=\Omega +\chi^{(0)}$, with the convention
that for even kernels the Euler Chern--Simons form is set to zero. 

\section{A new identity \label{eq:new_id}}

In this section we illustrate the new identity 
\be
\label{eq:fundamental_id} 
d\left(J_iK_j\right)= J_{ij}\,\chi_{ij}, \quad M_i\not\subset M_j,
\ee
where it is understood that the Euler--form of an empty normal bundle
is unity, $\chi(\emptyset)=1$. Its proof is worked out in the appendix. 
When $i+j<D$ \eref{eq:fundamental_id} is an identity between forms whose
degree exceeds $D$. In that case $M_i$ and $M_j$ have to be extended to
worldvolumes in a larger space--time, keeping the degrees of the $K$'s  and 
the $J$'s unchanged; see e.g. \cite{Lechner_Marchetti}.  

The ``non--extremality'' condition $M_i\not\subset M_j$ is needed to 
guarantee that the product  $J_iK_j$ is well--defined, implying 
that also its differential is so.  In the appendix it is then shown that 
$d(J_iK_j)$ is 1) closed, 2) invariant, 3) supported on $M_{ij}$ and  
4) constructed from the curvatures of $N_{ij}$. The above identity follows then
essentially for uniqueness reasons.

To simplify some formulae of the following sections, and motivated by
\eref{eq:fundamental_id}, we {\it define} for arbitrary intersections
$$
\left(J_iJ_j\right)_{reg}\equiv J_{ij}\,\chi_{ij}.
$$

Consider now an extremal intersection $M_i\subset M_j$, where 
the product $J_iK_j$ is ill--defined. Fortunately, as we will see,
in the dynamics of intersecting $D$--branes such a product will never 
show up. However, for notational covenience it will be useful to {\it define} 
\be 
(J_i K_j)_{reg} \equiv \left\{\begin{array}{ll} 
    J_i\,K_j & \mbox{if}\quad M_i \not\subset M_j,\\
    J_i\,\chi_j^{(0)} & \mbox{if} \quad M_i \subset M_j,\quad K_j
                       \quad\mbox{odd}\\  
    0                & \mbox{if} \quad M_i \subset M_j,\quad K_j
                       \quad\mbox{even.}
                      \end{array} \right. 
\label{reg}
\ee  
This definition is motivated as follows. If $M_i\subset M_j$ then for the
normal bundles we have $N_j=N_{ij}$, and therefore for the Euler--forms
$\chi_j=\chi_{ij}$. If $K_j$ is of odd rank, then $J_j$ is even and 
$\chi_j\neq 0$; if $K_j$ is even, then $J_j$ is odd and $\chi_j =0$.  
This implies that with the above definitions we have {\it in any case}
$$
d\,(J_i K_j)_{reg}=(J_iJ_j)_{reg}.
$$

We conclude this section presenting an alternative, but equivalent, way of
writing the information contained in
\eref{eq:fundamental_id}. We may rewrite  its l.h.s. in terms of the 
restriction of $K_j$ to $M_i$,  
$J_iK_j=J_i \left(K_j|_{M_i}\right)$. Since this restriction is 
well--defined, in this form we can apply Leibnitz's rule to get
$$
J_i\,d \left(K_j|_{M_i}\right)=J_{ij}\,\chi_{ij}.
$$
Denoting the $\delta$--function supported Poincar\`e--dual of $M_{ij}$ 
w.r.t. $M_i$ with 
${\cal J}_{ij}$ -- this is a form on $M_i$ and not on target--space --
we have $J_{ij}=J_i\,{\cal J}_{ij}$. The target--space relation 
\eref{eq:fundamental_id} is then equivalent to the relation on $M_i$, 
\be 
	d \left( K_j|_{M_i} \right)={\cal J}_{ij}\,\chi_{ij}. 
\label{eq:fundamental_id2} 
\ee 
We can go one step further and observe that,
if the intersection is effectively non--transversal i.e. $\chi_{ij}\neq 1$,
then the above relation is equivalent to the existence of a form 
${\cal L}_{ij}$ on $M_i$ such that
\be
\label{further}
K_j|_{M_i}-{\cal J}_{ij}\,\chi_{ij}^{(0)}=d {\cal L}_{ij},
\ee
transforming under $Q$--transformations of $K_j$ and under normal
bundle transformations of $N_{ij}$ respectively as
$$
\delta {\cal L}_{ij} = Q_j|_{M_i},\quad \delta {\cal L}_{ij}
=-{\cal J}_{ij}\,\chi_{ij}^{(1)},
$$
apart from closed forms. 

If the intersection is extremal, $K_j|_{M_i}$ is not defined and according
to above one would rather consider the expression 
$\chi_j^{(0)}|_{M_i}-{\cal J}_{ij}\,\chi_{ij}^{(0)}$, which vanishes 
identically since ${\cal J}_{ij}=1$.
This suggests to {\it define}, for $M_i\subset M_j$, 
${\cal L}_{ij}=0$.

\section{Basic applications}

We present here two basic applications of the above identity, to 
dual pairs of branes and to self--dual branes. These cases enter
as main building blocks in the construction of the action for arbitrary 
intersections, given in the next section. These two examples illustrate the
role played by $Q$--invariance, which is fundamental also in the general
case. For the sake of clarity we ignore here all other couplings 
but the minimal ones. We restore now the brane charges $g_i$ and Newton's 
constant $G$, taken until now as $g_i=1$ and $G=1/2\pi$. 
  
\subsection{Electromagnetically dual pairs of branes} 

Suppose that a $RR$--field strength satisfies the Bianchi
identity and equation of motion 
\ba 
	d R &=& g_j \, J_j  , \label{eq:Bianchi_simple_1} \\ 
	d * R &=& g_i \, J_i , \label{eq:Bianchi_simple_2}
\ea 
where $J_j$ ($J_i$) is the current 
on the high--dimensional (low--dimensional) magnetic (electric) $D$--brane 
with wordvolume $M_j$, $(M_i)$ and charge $g_j$ $(g_i)$. $M_i$ and $M_j$
form an electromagnetically dual pair whose dimensionalities satisfy 
$i+j=8$. For the  self--dual $D3$--brane the eq. of motion has to be replaced
by $R=*R$, but we do not consider this case in the present section. 

To write an action for such a system one must first solve
\eref{eq:Bianchi_simple_1}, introducing a potential for $R$, and to do this
one must search for an antiderivative of $J_j$ or of $J_i$ (or of both).

There are two candidates for such antiderivatives. The standard one is 
a Dirac--brane $M_{j+1}$, i.e. a brane whose boundary is $M_j$, with 
$\delta$--function supported Poincar\`e dual, say $W_j$. Then one has
$$
J_j=dW_j.
$$
The same can be done for $J_i$. The $(9-j)$--form $W_j$ carries by
construction $\delta$--like singularities on $M_{j+1}$ and hence also 
on $M_j$. Since the Dirac--brane is unphysical one must eventually
ensure that it is unobservable. In this case one would solve  
the Bianchi identity \eref{eq:Bianchi_simple_1} 
through $R=dC+g_jW_j$, and $C$ would carry the known singularities
along the Dirac--brane and on $M_j$

The second candidate for an antiderivative is a Chern--kernel $K_j$, which 
carries {\it invariant inverse--power like singularities on $M_j$},
$$
J_j=dK_j,
$$
and due to this fact the potential $C$ introduced according to 
$R=dC+g_jK_j$ is {\it regular} on $M_j$, because all singularities
are contained in $K_j$.
Since the Chern--kernel is defined 
modulo $Q$--transformations, one must eventually ensure that the theory
is $Q$--invariant, i.e. independent of the particular kernel one has 
chosen. 

We recall now the recipe developed in \cite{Lechner_Marchetti}
for writing an action
for the system \eref{eq:Bianchi_simple_1}, \eref{eq:Bianchi_simple_2}
if $M_i$ and $M_j$ have a transversal intersection. Then we will present 
its adaptation to a non--transversal one. (Remember that for a dual pair
of branes a transversal intersection amounts to no intersection at all, while 
a non--transversal one means simply $M_{ij}\neq\emptyset$).
The recipe goes as follows. Introduce a Chern--kernel for the 
magnetic brane
$J_j=dK_j$, and a Dirac--brane for the electric brane $J_i=dW_i$.
Then solve the Bianchi--identity \eref{eq:Bianchi_simple_1} according to
\be
\label{defR}
R=dC+g_jK_j.
\ee
Under $Q$--transformations of $K_j$ the potential must now also transform,
\be
\label{trans}
C^\prime=C-g_jQ_j,\quad K^\prime_j=K_j+dQ_j,  
\ee
to keep the field--strength $R$ invariant. For the restrictions on $M_j$
\eref{eq:Q_transform2} implies then, for an odd kernel, 
\be
\label{transpull}
\delta C|_{M_j}=-g_j\chi_j^{(1)},
\ee
while for an even kernel $C|_{M_j}$ is invariant. 
From these transformations one sees that the potential $C$ is a field regular
on $M_j$, because $\delta C|_{M_j}$ is finite. Equivalently, all singularities
of $R$ are contained in the Chern--kernel $K_j$, more precisely 
in the invariant form $\Omega_j$.  

The action, generating the 
equation of motion  \eref{eq:Bianchi_simple_2}, is given by 
\be 
S = \frac{1}{G} \int \left(\frac{1}{2} R * R +g_i\,R\,W_i \right).
 \label{eq:action_simple} 
\ee 
Under a change of the 
(unphysical) electric Dirac--brane, $W_i\rightarrow W_i+dZ_i$, this action 
changes by an integer multiple of $2\pi$, if the quantization condition 
$$
{g_ig_j\over G}=2\pi n
$$
holds, see e.g. \cite{Lech_Mar}. Moreover, the action $S$ is 
trivially free from gravitational anomalies. 

Two comments are in order. First, the eq. of motion 
\eref{eq:Bianchi_simple_2} could also be
obtained more simply from the action 
$ \frac{1}{G} \int \left(\frac{1}{2} 
R * R +g_i\,dC\,W_i \right)= \frac{1}{2G} \int R*R
+\frac{g_i}{G}\int_{M_i}C$. 
However, this action is inconsistent in that it
breaks $Q$--invariance in the bulk, because $C$ is not $Q$--invariant.
Second, introducing an eleven--dimensional manifold
which bounds target--space, the above $WZ$--term  can be rewritten 
equivalently as 
\be
\label{alter}
{g_i\over G}\int R W_i = {g_i\over G}
\int_{M_{11}}\left(RJ_i-g_jJ_jW_i\right)
={g_i\over G}\int_{M_{11}} RJ_i  \quad {\rm mod}\quad 2\pi,  
\ee
where in the last step we used that $\int J_j W_i$ is integer 
\cite{Lech_Mar}. The eleven--form ${g_i\over G}RJ_i$ is indeed 
closed modulo a well--defined integer form\footnote{``Integer forms'' are 
by definition forms
that integrate over an arbitrary manifold (closed or open) to an integer.
It can be shown that all such forms are necessarily $\delta$--functions on 
some manifold $M$. In particular our currents $J$ are integer forms, and 
a product of integer forms, whenever it is well--defined, is an integer 
form as well.}, because we are assuming that $M_i$ and $M_j$ are not
intersecting: $d({g_i\over G}RJ_i)=2\pi nJ_jJ_i$.

Let us now adapt this recipe to a dual pair with a non empty intersection
manifold $M_{ij}$, starting from \eref{alter}. 
In this case, due to the identity \eref{eq:fundamental_id},
the form ${g_i\over G}RJ_i$ is no longer 
closed, not even modulo integer forms, and there is the additional problem
that for extremal intersections the product $K_jJ_i$ is ill--defined. 
But, in turn, thanks to this identity we know how to amend the $WZ$--term.
Replace 
\be
\label{universal}
{g_i\over G}\int_{M_{11}} RJ_i \rightarrow
S_{WZ}\equiv {g_i\over G}\int_{M_{11}}\left((RJ_i)_{reg}-g_jJ_{ij}
\chi_{ij}^{(0)}\right),
\ee
where the subscript {\it reg} refers to the product $K_jJ_i$, defined in 
\eref{reg}: the integrand is now again a {\it well--defined closed} 
eleven--form.
This is the $WZ$--term anticipated in the introduction, see 
\eref{6a}, holding now for extremal intersections as well. 
 
For an extremal intersection,
$M_i\subset M_j$, due to $\chi_j=\chi_{ij}$ and $J_i=J_{ij}$,
the above $WZ$ simplifies to   
$$
S_{WZ}={g_i\over G}\int_{M_{11}}dCJ_i={g_i\over G}\int_{M_i}C. 
$$
Despite the formal cancellation of the ``anomalous'' term 
$\chi_{ij}^{(0)}$, the anomaly itself has not disappeared. Indeed,
the potential $C$ transforms under a $Q$--transformation, and  
thanks to \eref{transpull}, since 
$M_{ij}=M_i\subset M_j$, we have
$$
\delta C|_{M_i}=-g_j\chi_j^{(1)}= -g_j\chi_{ij}^{(1)}.
$$
The anomaly polynomial supported on $M_{ij}$ is then  
$-2\pi n\chi_{ij}$ also for extremal intersections, and the action remains 
$Q$--invariant in the bulk. 

As we observed, for non empty (non extremal) intersections it is 
$Q$--invariance that forces to put in \eref{universal} 
the $Q$--invariant combination $RJ_i$ instead of the closed form $dCJ_i$. 
So it is eventually $Q$--invariance in the bulk that requires
the presence of the anomalous term $J_{ij}\,\chi_{ij}^{(0)}$, needed 
to get back a closed eleven--form in the $WZ$.
This interplay between $Q$--invariance and anomalies will be a guiding
principle also in our construction of an anomaly--free effective action
for an arbitrary  system of intersecting $D$--branes in the next section. 

We may then summarize the properties of $S_{WZ}$ in \eref{universal} as 
follows. 1) It is written
in terms of a potential $C$ that carries an anomalous transformation
law but that is regular on $M_j$. 
2) It gives rise to the equation of motion \eref{eq:Bianchi_simple_2}.
3) It exhibits no singularities.
4) For transversal intersections ($J_{ij}=0$) it is well--defined modulo 
$2\pi$, $Q$--invariant and anomaly free. 
5) For non--transversal intersections it is well--defined, it carries the
gravitational anomaly $-2\pi n\chi_{ij}$ supported on $M_{ij}$, and it
is $Q$--invariant in the bulk.  6) For empty
intersections the second term drops, because $J_{ij}$ is vanishing, and 
one gets back the $WZ$--term for transversal intersections. 

We conclude and summarize this subsection, giving a ten--dimensional 
representation of \eref{universal}. For transversal intersections
one introduces an electric Dirac--brane for $M_i$, $J_i=dW_i$, and uses 
that \eref{universal} reduces to \eref{alter}, while for non--transversal 
ones one uses \eref{further}. The results are: 
\be
S_{WZ}={g_i\over G}\int_{M_i}C+\left\{\begin{array}{ll}
       {g_ig_j\over G}\int K_jW_i & \quad\mbox{for transversal 
                                                    intersections,}\\
       {g_ig_j\over G}\int_{M_i}{\cal L}_{ij}&\quad\mbox{for non--transversal 
                                                   intersections.} 
                                     \end{array}\right.
\label{ten}
\ee
We recall that for extremal intersections ${\cal L}_{ij}$ is zero by 
definition. The eleven--dimensional representation \eref{universal} for 
$S_{WZ}$ is, however, {\it universal} in that it holds for arbitrary
intersections, transversal or not. 

\subsection{Self--dual branes \label{sect:self_dual}} 

As second example we consider a self--dual brane, i.e. a brane $M_j$ with 
$2j=D-2$, which is coupled
to a field--strength satisfying a self--duality condition rather than 
a Maxwell equation, 
$$
dR=g_jJ_j, \quad R=*R.
$$
The brane we will be interested in is clearly the self--dual $D3$--brane
in $IIB$--Sugra.
Strictly speaking, this case can be treated without using the identity
\eref{eq:fundamental_id}, so 
we recall simply the results of \cite{Lechner_Marchetti}. 
One solves the Bianchi identity as above, $R=dC+g_jK_j$, and using the 
covariant $PST$--approach \cite{PST} to deal with the self--duality condition, 
one can write the action as 
$$
S={1\over 4G}\int \left(R*R+f_4f_4\right)+{g_j\over 2G}\int_{M_j}C.
$$
The additional (overall) factor of 1/2 will be explained in the next section.
A part from this the $WZ$ appearing here coincides with the one of the
extremal intersection of a dual pair discussed above; a self--dual brane
amounts in some sense indeed to $M_i=M_j$, a special case of 
$M_i\subset M_j$, $i+j=D-2$. The potential $C$ transforms under $Q$ according 
to \eref{transpull}. This means that $S$ entails the anomaly
polynomial 
$$
-{g_j^2\over 2 G}\,\chi_j=-{2\pi n \over 2}\chi_j,
$$ 
supported on $M_j$.

Due to the presence of the $NS$ $B$--field and of the gauge--fields on 
the branes, the technical details of the anomaly cancellation mechanism 
of a generic 
system of intersecting $D$--branes in $IIB$--Sugra appear slightly
involved. For this reason we presented the new ingredients of the 
Chern--kernel approach, which are crucial for
the entire construction, separately and in some detail
in these first four sections. 

We conclude this more general part
by stressing again that the Chern--kernel approach, as we saw, allows
to write well--defined actions which entail no 
ambiguities or singularities, despite the dangerous short--distance
configurations represented by branes intersecting non--transversally.
Furthermore, this approach does not require any regularization (or 
smoothing) of the currents, a procedure that would immediately 
run into troubles with the unobservability of the Dirac--brane,
see \cite{Tonin,Lech_Mar}.

\section{Anomaly cancellation  for intersecting $IIB$ $D$--branes}
 
In this section we consider the full interacting system of $IIB$ supergravity 
and abelian $D$--branes (one for each dimensionality) with arbitrary 
intersections. We recall first  
the quantum anomalies of the system, including from the beginning the $NS$ 
two--form $B$ in the dynamics. Then we give the full set of Bianchi 
identities and 
equations of motion and solve them in terms of regular potentials, 
applying systematically the 
proposals of eqs. \eref{5b} and \eref{66}. Next we present the action, 
check it is well defined and show it cancels all the anomalies. Eventually we discuss 
Born-Infeld actions and equations of motion on the branes. 

\subsection{Anomalies and the $B$ field \label{sec:anomalies}}

On each $D$--brane with worldvolume $M_i$ the pullback of
the NS $2$--form $B$ couples to the abelian gauge field $A_i$ living on
$M_i$ through the invariant field strength
\be 
	h_i = B - 2\pi \alpha^\prime F_i,\qquad F_i=dA_i.\label{eq:brane_YM}  
\ee 
Under a gauge transformation   $\delta B = d\Lambda$, the
$U(1)$ potential transforms as 
\be 
	\delta A_i = \frac{1}{2\pi\alpha^\prime} \Lambda, 
	\label{eq:A_gauge_transf} 
\ee 
where pull backs are understood. 

The anomaly polynomial that describes the anomalies of the system
has been derived in \cite{cinesi} and is given by the $12$--form 
\be 
	P_{12} = -\pi \sum_{i,j} (-1)^{\frac{i}{2}} \, 
	e^{\g(h_j - h_i)}
	\sqrt{\frac{\Ah(T_i)}{\Ah(N_i)}} \,
	\sqrt{\frac{\Ah(T_j)}{\Ah(N_j)}} \, J_{ij} \, \chi_{ij},
          \label{eq:anom_poly} 
\ee 
where $\g \equiv \frac{1}{4 \pi^2 \alpha^\prime}$, $\Ah$ is the roof
genus and $N_i$ and $T_i$ are the normal and tangent bundles respectively.
Since $h_j-h_i= 2\pi\alpha^\prime(F_i-F_j)$ the polynomial $P_{12}$ is 
independent of $B$ and invariant under \eref{eq:A_gauge_transf}, as it should. 
The term in the sum is symmetric under exchange of $i$ with $j$, so
that on single branes (which can be considered as self--intersection) 
a factor of $\pi$ appears
since summation is on diagonal terms $i=j$, while for intersecting
branes one recovers a factor of $2\pi$. \eref{eq:anom_poly}
reproduces indeed the anomaly polynomial for both single and intersecting
$D$--branes. Notice that the anomaly
on the (self intersecting) $D(-1)$-- and $D1$--brane 
$(i=0,2)$ vanishes since the degree $(10-i)$ of the Euler form $\chi_{ii}$ 
exceeds $i+2$, while the anomalies on self
intersecting $D3$, $D5$, $D7$--branes are different from
zero. The case of the $D9$ is special in that the worldvolume theory
is the super Yang--Mills part of type I String Theory \cite{Polchinski} and
its anomaly is cancelled by the Green--Schwarz mechanism. We do not
include this contribution in our discussion since we consider it well
known. 
 
The anomalies that have not been dealt with both in
\cite{Green} and \cite{cinesi} are those for pairs of dual branes, 
$D(-1)$--$D7$, $D1$--$D5$, $D3$--$D3$, where the integrand in
\eref{eq:anom_poly} for dimensional reasons reduces entirely to the Euler 
form. In this case
it was not understood how to perform an inflow of charge and, as
mentioned in \cite{cinesi}, ``a more powerful approach is
needed''. This can be achieved using Chern--kernels as we have seen in
the previous sections.  

The charge of the $D$--brane with worldvolume $M_i$  is given by 
\cite{Polchinski} 
\be 
	g_i = \sqrt{2\pi G} \, \g^{\frac{i-4}{2}}, \quad
	g_i \, g_{8-i}=2\pi G.  
\label{cariche}
\ee
As noticed in \cite{cinesi,Green} a crucial step towards anomaly
cancellation is to notice that the anomaly polynomial 
\eref{eq:anom_poly} is partially 'factorized'. 
On each $D$--brane one can introduce the closed and invariant polynomial 
\be 
	Y_{i} = e^{-\frac{F_i}{2\pi}} \, \sqrt{\frac{\Ah(T_i)}{\Ah(N_i)}}
              =  d \, Y^{(0)}_{i} + 1 . 
	\label{eq:Y} 
\ee 
The forms $Y_i$ are those appearing in the Bianchi identities
and equations of motion of \cite{cinesi,Green}, where $B$ was kept zero.
For a non vanishing $B$--field one notices that \eref{eq:Y} is not
invariant under \eref{eq:A_gauge_transf}, so what should appear in the 
Bianchi identities and equations of motion is rather the invariant expression
\be 
\Yb_{i}= e^{\g h_i} \, 
	\sqrt{\frac{\Ah(T_i)}{\Ah(N_i)}} 
	= e^{\g B_i} \,Y_{i}. 
	\label{eq:Yb} 
\ee 
These $\Yb_i$ are, however, no longer closed and satisfy 
\be 
	d \Yb_i =  \g \Yb_i \, H, \qquad H=dB. \label{eq:d_Yb} 
\ee 
In terms of these forms one can define the following forms of
degree $n = 2,\dots, 10$, 
\be 
	\Delta_n =  g_{10-n}  \sum_{i}
	\, (-1)^{\frac{i}{2}} J_{i} \, \Yb_{i} \,  ,
	\label{eq:Delta}  
\ee 
where it is understood that on the r.h.s. one has to extract the $n$--form
part. \eref{cariche} implies then the important relation 
\be 
	d \Delta_n =  \Delta_{n-2} \, H.
	\label{eq:d_Delta} 
\ee
 
These forms are crucial since they allow eventually to factorize the
anomaly polynomial completely. Indeed, it is not difficult to show that 
one can rewrite  \eref{eq:anom_poly} as 
\be 
	P_{12} = -\frac{1}{2G} \sum_n (-1)^{n/2} \left( \Delta_n
	\Delta_{12-n}   \right)_{reg} , \label{eq:anom_poly2} 
\ee 
where the 'regularized' product of two currents has been defined in
the previous section. This formula holds in the case of non trasversal
intersections for all couples of dual branes. In case these intersections
are  all transversal one has to add the term $-2\pi (J_6 J_2 - J_8 J_0 )$
which subtracts the same term from the above expression. Notice however 
that, since for transversal intersections the form $J_6 J_2 - J_8 J_0$ is  
integer, it gives rise through the descent formalism to an anomaly which is 
a well defined integer multiple of $2\pi$, and can be disregarded.

Notice also that \eref{eq:anom_poly2} is independent on $B$ -- despite
its explicit
appearance -- as is obvious by construction. This can also be checked 
explicitly noticing that under a generic variation of B one has
\be 
	\delta \Delta_n= \Delta_{n-2}\,\delta B.
	\label{eq:delta_Delta} 
\ee 
$\delta P_{12}$ vanishes under such a variation
thanks to the alternating signs
$(-1)^{n/2}$ in eq. \eref{eq:anom_poly2}. This explains also the 
appearance of those signs.

 Evenutally, since $P_{12}$ is invariant under \eref{eq:A_gauge_transf}
 there exist a Chern--Simons form and a second descent which respect
 this symmetry, $P_{12}=d P_{11}$, $\delta P_{11}=dP_{10}$. The resulting 
 anomaly 
 \be 
	\mathcal{A} = \int P_{10} 
 \ee 
 can therefore always be choosen to respect this symmetry, too. 

\subsection{Bianchi identities and equations of motion} 

In this section we describe the theory of IIB supergravity in presence
of arbitrary $D$--branes and with a non zero $B$
field turned on. We give new Bianchi identities and equations of motion  
that apply in the presence of $D$--branes . Using the techniques developed in the previous
sections we solve Bianchi identities in terms of regular
potentials, and then proceed to write down a classical action that
gives the equations of motion. $Q$--invariance of this classical
action generates an anomalous transformation law that exactly cancels
the quantum anomaly. 
 
Start defining the formal sum of
generalized currents as 
\be 
	\Delta = \sum_{n}  \Delta_n , 
\ee 
where $\Delta_n$ is defined in \eref{eq:Delta}. The full set of
Bianchi identities and equations of motion is given by the compact
formula 
\be 
	d R = R H + \Delta ,  
	\label{eq:Bianchi_id} 
\ee 
where $R_9 = * R_1$, $R_7 = * R_3$ and $R_5 = * R_5$. 
Three comments are in order. First, in the limit where each brane
charge $g_i$ is set to zero, \eref{eq:Bianchi_id} reduces to the
Bianchi identities and equations of motion of free $IIB$
supergravity.  
Secondly, \eref{eq:Bianchi_id} is well defined since the
right hand side is a closed form, as can be seen using
\eref{eq:d_Delta}. Third point to mention is electro-magnetic duality. 
While an equation of the form $d R = J$ is electro-magnetically  symmetric, 
the full equation \eref{eq:Bianchi_id} is not, because of the $R H$ and $J \overline{Y}$ 
terms. These latter can be thought of as currents
associated to smeared branes.  

The compact formula describing solutions of the Bianchi identities is 
\be 
	R = dC - C H +   f ,
	\label{eq:Bianchi_sol}  
\ee 
where $f_n$, $n=1,3,5$, is the $n$--form 
\be 
	f_n=  g_{9-n}  \sum_{i}
	(-1)^{\frac{i}{2}} K_i \Yb_i  \label{eq:f} , 
\ee 
and it satisfies 
\be 
	df = f H + \Delta . 
\ee 
Notice that under $Q$ transformations $\delta K_i = dQ_i$ and  the RR
curvatures are invariant provided the 
potentials transform as 
\be 
\delta C_n = - g_{8-n} \sum_{i} (-1)^{i/2} Q_i \, \Yb_i ,  
\ee 
and therefore the pullback of the potentials on the branes is
regular. Such pullback amounts to an anomalous transformation of the
potential, which plays an important role in cancellation of
anomalies.

As a specific example, the corrected form of the Bianchi identities
for $R_1$ and $R_3$ is  
\ba 
	d R_1 &=&  g_8 J_8 , \label{eq:Bianchi_id_R1} \\  
	d R_3 &=& R_1 H + g_6 ( - J_6 + J_8 \Yb_{8,2} )  ,  
	\label{eq:Bianchi_id_R3}      
\ea 
(by $\Yb_{8,2}$ we mean the degree 2 part of the $\Yb$ form defined on
the $D7$ brane). 
Now we introduce Chern-kernels $K_8$, $K_6$ (and
$K_4$ for $R_5$) associated to the $D7$, $D5$ (and $D3$) branes that appear in the
Bianchi identities. Further sources appearing in the equations of
motion, $J_2$ and $J_0$, have instead to be treated using Dirac branes
$W_2$ and $W_0$, as explained in \cite{Lechner_Marchetti}. 
The solution we propose for the Bianchi identities is  
\ba 
	R_1 &=& d C_0 + g_8 K_8 , \label{eq:R1} \\ 
	R_3 &=& d C_2 - C_0 H + g_6 ( - K_6 + K_8 \Yb_{8,2} ) , 
	\label{eq:R3}     
\ea 
and similarly for $R_5$. 
It is straightforward to check that these definitions ensure that the
Bianchi identities are satisfied, using \eref{eq:d_Yb},
\eref{cariche}. Notice the fact that this solution requires the $\Yb$ forms
to be extended to target space forms. One could object that a more standard way to solve
Bianchi identites that involves only $Y$ forms evaluated on branes
would rather be 
\ba 
	R_1 &=& d \tilde{C}_0 + g_8 K_8 , \\ 
	R_3 &=& d \tilde{C}_2 - \tilde{C}_0 H + g_6 ( - K_6 + J_8
	Y^{(0)}_{8,1} + K_8 B ) \label{eq:wrong_R3}   , 	
\ea 
and similarly for $R_5$. This is the approach used in \cite{cinesi},
and we remark
that there the solution of Bianchi identities is
incomplete, in that it misses the terms were 
$K$ appears without any $Y$ form.   
However, this definition leads to singular potentials and therefore to inconsistencies. 
Consider for
example an anomalous gauge transformation of $Y^{(0)}_{8,1}$ in
\eref{eq:wrong_R3}, $\delta Y^{(0)}_{8,1} = d Y^{(1)}_{8,0}$. Since
the curvature $R_3$ is invariant the potential $\tilde{C}_2$ 
transforms accordingly as 
\be 
	\delta \tilde{C}_2 = - g_6 J_8 Y^{(1)}_{8,0} . 
\ee 
The variation of
$\tilde{C}_2$ is always
singular on the $D7$--brane so that 
$\tilde{C}_2$ itself is singular. Similarly, one shows that an analogous $\tilde{C}_4$ would 
be singular on the $D7$
and $D5$--branes.

We now show that the RR curvatures are independent of the extension 
of the $\overline{Y}$ forms. Consider target space forms
$Y_i= dY^{(0)}_i + 1$. Under a change of extension they transform as 
\be 
	\delta Y_i = d X_i , \label{eq:Y_change} 
\ee 
where $X_i$ is an arbitrary target space form such that  
\be 
	J_i X_i = J_i \left. X_i\right|_{M_{i}} = 0 , 
\ee 
since $Y_i$ is well defined on its $D$--brane. From here on we will
refer to these as $X$--transformations. Here we explicitly consider the $R_3$
curvature, but similar formulae apply to all the potentials $C$. 
Under \eref{eq:Y_change} $R_3$ is invariant if $C_2$ shifts by 
\be
	\delta C_2 = - g_6 K_8 X_{8,1} .  \label{eq:C2_transf}
\ee 
Such $X$--transformations of the
potentials always exist due to consistency of the
Bianchi identities.  
Even though $K_8$ does not admit limit close to the $D7$--brane, 
it remains finite and the product $K_8 X_{8,1}$ is well
behaved and goes to zero since $X_{8,1}$ goes to zero. 
This proves that the RR potentials as
defined by \eref{eq:Bianchi_sol} are completely regular close to the
branes, and that the curvatures do not depend on the arbitrary
extensions of the $Y$ forms. In the next section we will present the
action for the system and see that it does not depend on such extensions.

\subsection{Action and anomaly cancellation \label{sec:action}}

In this section we present an action that gives rise to the equations
of motion \eref{eq:Bianchi_id}. 
There are two ways to discuss the Wess-Zumino part of the action. One
possibility is write it as the integral of a
closed $11$--form. The advantage of this formulation is that it is 
the most clear one: it immediately displays $Q$ and $X $--invariances of the theory 
and how the gravitational anomaly is
cancelled. However, even if the procedure is rigorous and does not depend on the
arbitrary extra dimension, nevertheless it is a natural expectation to
ask for the existence of a well defined ten-dimensional action. Our
approach to the problem will be that of presenting at first the Wess-Zumino 
part of the action
as the integral of an $11$--form and discuss its properties. A well defined $10D$ action
will be given in the end of the section, and its relationship with the former discussed in
the appendix. 
 
We define the total effective action of the theory as the sum of a
classical and quantum part 
\be 
	\Gamma = \frac{1}{G} \left( S_{kin} + S_{WZ} \right) +
	\Gamma_{quant} , \label{eq:action} 
\ee  
where $\Gamma_{quant}$ generates the anomalies described in
sect.\ref{sec:anomalies}. $S_{kin}$ is given by 
\be 
	\begin{array}{lcl} 
	S_{kin} &=& \int d^{10} x \sqrt{-g} e^{-2\phi}
	R + \frac{1}{2} \int \, e^{-2\phi} \left( 8 H_1 * H_1 + H_3 * H_3
	\right)  \\ 
	&+&  \frac{1}{2} \int \, \left( R_1* R_1 - R_3 * R_3 +
	\frac{1}{2} R_5 * R_5 - \frac{1}{2} f_4 * f_4 
	\right) ,  
	\end{array} \label{eq:S_kin} 
\ee 
where the field $f_4 = \iota_v (R_5 - * R_5)$ appears as part of the PST
approach \cite{PST}, enforcing the self-duality equation, and $v^m$
is the unit vector            
\be 
	v_m \equiv \frac{\partial_m a}{\sqrt{\partial a \partial a}},
	\;\;\; v_m v^m = 1 , \label{eq:PST_vector} 
\ee 
where $a$ is an arbitrary scalar field. The PST formulation then
guarantees that $a$ is a non propagating, auxiliary field. 

To describe the Wess-Zumino term instead take an
eleven dimensional manifold $M_{11}$ whose boundary is the spacetime
$M_{10}$ of the theory, $\partial M_{11} = M_{10}$. Assume no
topological obstruction, and take the extra dimension to be parallel
to the $D$--branes so that the degree of $J$ forms, which counts the
number of normal directions to a brane, is not changed. Then one can
write the Wess-Zumino as 
\ba 
	S_{WZ} &=& \int_{M_{11}} L_{11} , \\ 
	d L_{11} &=& 0 ,  
\ea 
and $L_{11}$ is given by 
\be 
	L_{11} =  \left[ - \frac{1}{2} R_5 \left( R_3 H + \Delta_6
	\right) - R_3 \Delta_8 - R_1 \Delta_{10} \right]_{reg} - G
	P_{11}  
	. \label{eq:WZ1} 
\ee 
$P_{11}$ is the Chern-Simons form of the anomaly and is given by 
\be 
	P_{11} =  \pi \sum_{i,j} (-1)^{i/2+1} P_{ij}^{(0)}
	J_{i j} \chi_{ij} + 2\pi ( 
 J_{2-6}
	\chi_{2-6}^{(0)} - J_{0-8} \chi_{0-8}^{(0)} ) - \pi J_4 \chi_4^{(0)}     , 
	\label{eq:P11} 
\ee 
and $P_{ij}^{(0)}$, in turn, is a Chern-Simons form 
of\footnote{Notice that $P_{ij} = dP_{ij}^{(0)} +1$, which explains
  the form of \eref{eq:P11}.} 
\be 
	P_{ij} =  e^{\frac{\left( F_i - F_j \right)}{2\pi}} 
	\sqrt{\frac{\Ah(T_i)}{\Ah(N_i)}} \,
	\sqrt{\frac{\Ah(T_j)}{\Ah(N_j)}} .  
      \label{eq:Pij} 
\ee 
Some comments are in order. A direct check shows that the classical
part of the action 
gives the correct equations of motion. Moreover, $Q$ and $X$--invariances in
this picture are immediately displayed, since only RR invariant
curvatures appear. 

Anomaly
cancellation is guaranteed since the only term in the action which is
not invariant under anomalous transformations is $- \int P_{11}$,
whose variation exactly cancels the quantum anomaly. Lastly, notice
the factor of $1/2$ in the minimal coupling $R_5 \Delta_6$ of
eq.\eref{eq:WZ1}. On one side it is an artifact of the PST formalism (see the
kinetic part of the action), and should not misunderstood as a novel
feature. On the other side, it combines exactly with $\pi J_4 \chi_4^{(0)}$ leaving 
only a $C_4$ potential term, that cancels the anomaly on the self-dual $D3$--brane 
according to what explained in sect.\ref{sect:self_dual}. 

We now write down the Wess-Zumino term in a
ten dimensional fashion. Here we simply display it as it is,
leaving the proof of its equivalence with \eref{eq:WZ1} to the
appendix. We split the Wess-Zumino into a term depending on the
potentials and a remainder:  
\be 
	S_{WZ} = \int ( L_{10}^{C} + L_{10}^{rem} )
	. \label{eq:WZ_10D} 
\ee 
The part depending on the potentials is 
\be 
	L_{10}^{C} = -\frac{1}{2} C_4 \left( R_3 H +\Delta_6  \right)
	- \frac{1}{2} C_2 H f_5 - C_2 \Delta_8 - C_0 \Delta_{10} .  
	\label{eq:L10_C} 
\ee 
The remainder can be written in two ways. In the case of transversal
intersection for dual branes it is given by 
\be 
	L_{10}^{rem} = \frac{1}{2} \left[ f_3 f_7 - f_1 f_9 + 2\pi G
\sum_{i,j} (-1)^{i/2} P_{ij}^{(0)} K_i J_j + 4\pi G ( K_6 W_2 - K_8 W_0
	) \right] , \label{eq:L10_rem} 
\ee 
where the forms $f_7$, $f_9$ are
formally defined as in \eref{eq:f}, but using Dirac branes $W_2$ and $W_0$ instead
of $K_2$ and $K_0$. For non-transversal intersection the two
last $K_i W_j$ terms have to be modified according to eq.\eref{ten}.  

In this $10D$ picture, $Q$ and $X$--invariances
are hidden and have to be checked one by one. Anomaly cancellation
arises for non dual branes from the terms $P_{ij}^{(0)} J_i K_j$. 
For dual branes with non extremal intersection it arises from the terms of the kind
$\mathcal{L}_{ij}J_i$. If the intersection is extremal then
$\mathcal{L}_{ij}=0$ and the anomaly is cancelled by the anomalous
transformation of the potentials. This is always true for the
$D3$--brane. All the other
$Q$--variations instead are cancelled between terms in $L_{10}^C$ and
in $L_{11}^{rem}$.

\subsection{Born-Infeld actions and equations of motion on the branes
\label{sec:BI}}
 
In this section we describe the dynamics of $U(1)$ fields on each
$D$--brane and of the NS form $B$. The action \eref{eq:action}
describes all the dynamics of RR fields but, as it stands, is not
complete. One has to add to it Born-Infeld terms for the $U(1)$ fields on
each brane: 
\be 
	\Gamma = \frac{1}{G} \left( S_{kin} + S_{WZ} +  S_{BI} \right) +
	\Gamma_{quant} , \label{eq:action2} 
\ee  
with 
\ba 
	S_{BI} &=& \sum_i g_i I^i_{BI}, \\ 
	I^i_{BI} &=& - \int_{Di} dx^{i} \, e^{-\phi} \sqrt{-det \left(
	g^i_{mn} + B_{mn} -  2\pi \alpha^\prime F^i_{mn} \right)} . 
\ea 
From such Born-Infeld term on can define generalized field strenghts 
\be 
	\htw^{i}_{mn} = \frac{2}{\sqrt{-det g^i_{mn}}} \, \frac{\delta
	I^i_{BI}}{\delta B_{mn}} , 
\ee 
justified by the fact that under a variation of the field $B$ one
has $\delta I^i_{BI} = \int_{Di} \delta B * \htw^{i}$. In terms of these
one can obtain, after a straightforward but lenghty calculation, the
equation of motion of $B$: 
\be 
	d * H = R_3 R_5 - R_1 R_7 + \sum_i g_i J_i * \htw^i ,
	\label{eq:B2_eom} 
\ee 
and the equations of motion for the $U(1)$ field strenghts, that we
report in appendix. 
These new equations of motion have three important properties. First of
all, they are invariant under $Q$ and $P$--transformations. This is
required by consistency since the action we wrote down is invariant in
first instance. $Q$--invariance happens because a direct check shows
that such equations display no dependence on Chern-kernels
$K_i$. $P$--invariance is evident since the equations are expressed in
terms of $\Yb$ forms pulled back on the appropriate
branes and of RR curvatures. Second point is that the equations are explicitly invariant under gauge
transformations of $B$ 
\ba 
	\delta B &=& d \Lambda , \\ 
	\delta A^i &=&  \frac{1}{2\pi\alpha^\prime} \Lambda|_{Di}
	,  
\ea 
again as it should, by consistency. Third property is that, as expected, the $U(1)$ 
theory on the branes is anomalous. If one writes the equations of
motion in the form 
\be 
	d * \htw^i = \tilde{j}_i , \label{eq:d*h} 
\ee 
then an explicit check shows that 
\be 
	d \tilde{j}_i \neq 0 . \label{eq:dj_non_zero} 
\ee 
However the equation of motion for $B$ \eref{eq:B2_eom}, even though it contains the anomalous
field strenghts $\htw^i$, is
non anomalous.  
 
The action constructed so far only involves $C_0$, $C_2$, $C_4$ potentials. 
Often in the literature the action is written in term of all the RR potentials, and hence  
in the next section we rewrite our results in a duality-invariant language.

\section{PST duality--invariant formulation \label{sec:dual_version}}
 
In order to make contact with the formulation of \cite{cinesi} in this
section we construct the action for the same system but using all the
possible RR potentials $C_i, i= 0, 2, \dots 8$, instead of the minimal
ones $C_0$, $C_2$, $C_4$. A duality-invariant formulation may also be
useful for the purpose of analysing the flux quantization of dual
potentials, or for dimensional reductions involving dual branes and
dual potentials. 
 
Let us then introduce the new RR potentials $C_6$, $C_8$ and define the
forms $f_7$ and $f_9$ as in \eref{eq:f}, but this time using proper
Chern-kernels $K_2$ and $K_0$ instead of Dirac branes. Introduce RR
curvature $R_7$ and $R_9$ using the same recipe of
\eref{eq:Bianchi_sol}. In order to deal with $C_6$ and $C_8$ one has
to exploit the PST formalism. Introduce an arbitrary scalar field $a$
and 
and construct the unit vector $v^m$ as in \eref{eq:PST_vector}. 
 In term of $v^m$ construct the forms 
\ba 
	r_0 &\equiv & \iota_v ( R_1 - * R_9 ) , \\  
	r_2 &\equiv & \iota_v ( R_3 - * R_7 ) . 
\ea 
Then the PST duality-invariant action is 
\be 
	S_{dual} = S_{kin} + S_{WZ} +  S_{BI} + \frac{1}{2} \left( r_2 *
	r_2 - r_0 * r_0 \right) . \label{eq:dual_action}
\ee 
Such action has all the PST symmetries necessary to prove that $a$ is
an auxiliary field and that the conditions $R_9 = * R_1$, $R_7 = *
R_3$ are enforced (see \cite{PST}, \cite{Lechner_Marchetti}). In
order to make contact with the usual formulations one can use the
following identities: 
\ba 
	R_3 * R_3 - r_2 * r_2  &=& (R_3 , R_7) \, P(v) \left(
	\begin{array}{l} R_3 \\ R_7 \end{array} \right) - R_3 R_7 , \\ 
        R_1 * R_1 - r_0 * r_0  &=& (R_1 , R_9) \, P(v) \left(
	\begin{array}{l} R_1 \\ R_9 \end{array} \right) + R_1 R_9 , 
\ea 
where $P(v)$ is the operator valued matrix 
\be 
	\left( \begin{array}{cc} 
		  - v \iota_v * & v \iota_v \\ 
		  v \iota_v     & - v \iota_v * 
	       \end{array} \right) . 
\ee 
Substituting this into the action \eref{eq:dual_action} gives 
$S_{dual} = S_{kin,dual} + S_{WZ,dual} + S_{BI}$, where $S_{kin,dual}$ is 
a kinetic term for all the
RR potentials given by 
\be 
	S_{kin,dual} = \frac{1}{2} \int \, \left[ (R_1 , R_9) \, P(v) \left(
	\begin{array}{l} R_1 \\ R_9 \end{array} \right) - (R_3 , R_7) \, P(v) \left(
	\begin{array}{l} R_3 \\ R_7 \end{array} \right) + \frac{1}{2}
	R_5 * R_5 - \frac{1}{2} f_4 * f_4 \right]  ,  
	 \label{eq:S_kin_dual} 
\ee 
where there is symmetry under $R_1 \leftrightarrow R_9$, $R_3
\leftrightarrow R_7$, see \cite{Lech_Mar}. $S_{WZ,dual} = \int L_{dual}$ is 
a modified
Wess-Zumino term whose $11D$ version reads  
\be 
	L_{11,dual} = - \frac{1}{2} \sum_{n} R_{n+1}
	\Delta_{10-n} - G P_{11}   , 
\ee 
while the $10D$ one is 
\be 
	\begin{array}{lcl} 
	L_{10,dual} &=& -\frac{1}{2} \sum_{n} C_n
	\Delta_{10-nn} + \frac{2\pi}{2} G \left[ \sum_{i,j} (-1)^{i/2}
	P_{ij}^{(0)} ( K_i J_j )_{reg} + 2 ( K_6 W_2 - K_8 W_0 ) \right] \\ 
	&=& -\frac{1}{2} \sum_{n} R_{n+1} 
	f_{9-n} + \frac{2\pi}{2} G \left[ \sum_{i,j} (-1)^{i/2}
	P_{ij}^{(0)} ( K_i J_j )_{reg} + 2 ( K_6 W_2 - K_8 W_0 ) \right] . 
	\end{array} \label{eq:L10_dual} 
\ee 
Again the usual remark for non-trasversal intersections of dual branes
applies, where eq.\eref{ten}  should be used. 
Now we can try to make contact with eq.(2.11) of \cite{cinesi}. In our
notation it says that on each brane the Wess-Zumino term goes like 
\be 
   -\frac{1}{2}\int_{M_i}\left( \tilde{C}_i + R\, Y_i^{(0)} \right) . 
  \label{eq:cinesi} 
\ee
From
eq.\eref{eq:f} one
can decompose the forms $f_n$, in the limit $B\equiv 0$, as 
\be 
	f_n =  g_{9-n} \sum_i 
	(-1)^{\frac{i}{2}}   \left[  ( J_i Y_i^{(0)} + K_i ) 
	+ d( K_i  Y_i^{(0)})  \right]_n , \label{eq:decomp} 
\ee 
and plug them into the second line of \eref{eq:L10_dual}. 
Consider the first three terms in \eref{eq:decomp}. $f_n$ on
its own has only inverse power singularities near each brane, but the
first and third term in the decomposition individually display $\delta$--like
singularities. Therefore, in \eref{eq:L10_dual}  it is not allowed to multiply each single
term times a RR curvature, but only the whole sum.  
Suppose however we want to 
formally forget about this difficulty. Then 
we can see that the first term in \eref{eq:decomp} reproduces 
the second term of \eref{eq:cinesi}.   
The second term has two effects. Part of it is multiplied in
eq.\eref{eq:L10_dual} times the potential
part in $R$. Joint with some of the $P_{ij}^{(0)}(K_iJ_j)_{reg}$ terms, it
reconstructs the $\tilde{C}_i$ part of \eref{eq:cinesi}. Remember that
the passage from $C_i$ to $\tilde{C}_i$ is purely formal since the latter
is ill-defined. The remaining part of the second term in
\eref{eq:decomp}, together with the third term $d(K_i
Y_i^{(0)})$, are dependent on the Chern-kernels
$K_i$ and cancel completely with the rest of the
$P_{ij}^{(0)}(K_iJ_j)_{reg}$ terms 
of \eref{eq:L10_dual}. This is guaranteed since the (ill-defined)
potentials $\tilde{C}_i$ of \cite{cinesi} are $Q$-invariant and so the
$K$-dependent terms have to disappear.

In conclusion, in the formal approximation when one
can forget $\delta$--like divergencies, in the limit $B \equiv 0$ and
assuming it is possible to use the $\tilde{C}_i$ potentials, one
exactly recovers the Wess-Zumino term of \cite{cinesi}, plus the extra
terms that cancel the anomaly for dual branes. The anomaly
cancellation for the self-dual $D3$--brane is given by transformations
of $C_4$ which are present only in the Chern-kernel formulation and
cannot be reproduced in the context of \cite{cinesi}.

\section{Conclusions and outlooks \label{sec:conclusions}}
 
We conclude by summarizing our results and commenting on their
extension. 
 
We have considered the system $IIB$ supergravity
interacting with all possible combinations of single $D$--branes, 
with arbitrary intersections as long as there are no topological 
obstructions. We have constructed a regular action that gives the equations of motion,
which is written in terms of potentials that are everywhere well defined. 
 
We have provided a correct understanding of the mechanism of charge
inflow using Chern-kernel techniques. In particular we have shown
that, for pairs of dual branes which had proven to be intractable
before, charge inflow is not produced by curvatures but either by potentials, in the case of 
the self-dual $D3$--brane and extremal intersections of dual branes, 
or by the $\mathcal{L}$ forms for non-extremal, non-transversal intersections 
of dual branes. Another important part of the understanding of charge
inflow, that is used in order to implement a Chern-kernel analysis, is
the new fundamental identity \eref{eq:fundamental_id}.  

The Wess-Zumino term we obtained differs from other expressions 
that appeared in the literature, like those of \cite{Green},
\cite{cinesi}, \cite{ScruccaSerone}
 and in particular it contains extra terms which
are related to anomaly cancellation for pairs of dual
branes. Moreover, it contains all the corrections due to the presence
of the NS form $B$. 
 
We have obtained the full, corrected equations of motion
for $B$ interacting with RR fields, gravity and $U(1)$ Yang-Mills
fields, and for the $U(1)$ fields themselves. The classical $U(1)$
theory on the branes is anomalous but quantum corrections restore the full symmetry. 
 
We insist on remarking that Chern-kernel techniques have wide
application to all theories with extended objects, and not only
$D$--branes of supergravity. They can be used for example to deal with
orientifolds, like in problems considered in \cite{Diaconescu,Lust},
with $O$--planes \cite{Scrucca_O_planes}, with non BPS branes
\cite{Scrucca_non_BPS}. Another possible system to which apply these
techniques is $IIA$ supergravity in presence of $D$--branes. There,
branes have odd dimensional worldvolume but they still admit anomalies
on their intersections. 
 
Another possible generalization is to couple our system of $D$--branes
to an $NS5$--brane, which is interesting since in that case the NS
curvature would not be closed. Its treatment should go along the lines
of \cite{NS5} and we expect it to be straightforward to implement. 
 
Lastly, we discuss generalization to the $U(N)$ case. In this case, it
is reasonable to argue that, in costructing physical $U(N)$ fields on
each brane, the colourless NS form $B$ will be coupled to some
$U(1)$ subgroup on $U(N)$. Let $\mathcal{F}$ then be the full $U(N)$
curvature, and decompose it into a $U(1)$ part $F$, that couples to
$B$ as in \eref{eq:brane_YM}, and an $SU(N)$ part with curvature
$\tilde{F}$. Since the $U(1)$ part commutes with the rest, it is easy
to see that the non-abelian Chern character that enters in the anomaly 
has to be generalized to 
\be 
	ch \left( \frac{\mathcal{F}}{2\pi} \right) \rightarrow 
	e^{- \gamma h} ch \left( \frac{\tilde{F}}{2\pi} \right) . 
\ee 
This would be the ingredient necessary to form the new $\Yb$
forms. Since $B$ enters in the $U(1)$ part the identity
\eref{eq:d_Yb} continues to hold and from this one is able to impose
again Bianchi identities, equations of motion, and to find a
Wess-Zumino term for the action from which they come from. The only
limitation is that one does not have a Born-Infeld action that is
uniquely fixed so far, and therefore for the time being it is not possible to find
equations of motion for $B$ and the Yang-Mills fields on the brane.

\section{Appendix: proof of the new identity}

The following proof holds for arbitrary Chern--kernels, even or odd. 
We begin by considering the properties of $J_iK_j$. As we saw, 
$K_j=\Omega_j+\chi_j^{(0)}$ is singular on $M_j$ because 
$\Omega_j$ involves $\yh^a=y^a/|y|$, and $M_j$ stays at $y^a=0$.
But since $M_i\not\subset M_j$ the product
$$
J_iK_j= J_i \left(K_j|_{M_i}\right)
$$
is well--defined and, therefore, in the sense of distributions also
its differential is so. The subtle point is only that one can not apply
Leibnitz's rule to evaluate it, because  the product has 
(inverse--power and $\delta$--like) singularities on $M_{ij}$. Away from
$M_{ij}$ one can apply Leibnitz and there the result is $d(J_iK_j)=0$.
This means that  $d(J_iK_j)$ is supported on $M_{ij}$ and hence proportional
to $J_{ij}$,
$$
d\left(J_iK_j\right)=J_{ij}\,\Phi,
$$
for some form $\Phi$ defined on $M_{ij}$.
Furthermore, since the l.h.s. is closed also $\Phi$ must be a closed form. 
Moreover, $\Phi$ must be a completely invariant form
as is the l.h.s.,
because $J_i$ is intrinsically defined and $K_j$ transforms as
$K_j^\prime= K_j+ dQ_j$, where $Q_j$ is {\it regular} on $M_j$. This means 
that one can apply Leibnitz and $d(J_idQ_j)=0$. 

Furthermore, $\Phi$ can depend only on the curvature components of the 
intersection of the normal bundles $N_{ij}$. This can be seen 
as follows. Since $K_j$ is made out only of gravitational curvatures 
belonging to $N_j$, also $\Phi$ is a polynomial made out only of 
(a subset of) those curvatures. If (in addition to $M_i\not\subset M_j$) 
we have also 
$M_j\not\subset M_i$, we can apply Leibnitz to $d(K_iK_j)=K_iJ_j\pm J_iK_j$,
giving $d(J_iK_j)=\pm d(J_jK_i)$. This implies that $\Phi$ depends, 
moreover, only on the curvatures of $N_i$, and hence only on those of 
$N_{ij}$. If on the contrary $M_j\subset M_i$ then, using e.g. 
the regularizations of \cite{Lechner_Marchetti}, one can show that 
$d(K_iK_j)=\chi_i^{(0)}J_j\pm J_iK_j$. Applying the differential to this 
one gets directly \eref{eq:fundamental_id}, since in this case
$J_j =J_{ij}$ and $\chi_i=\chi_{ij}$. 

Eventually,
$\Phi$ is a form of degree $dim(M_{ij})+D-(i+j)$ and it is odd under
parity. $\Phi$ shares all these properties uniquely with  the Euler--form 
of $N_{ij}$ and we conclude therefore that it is proportional to it. 

A cohomological argument can finally be used to fix the proportionality 
coefficient to one. 
Perform a regularization $K_j\rightarrow  K_j^\varepsilon$,
$J_j\rightarrow  J_j^\varepsilon$, $J_j^\varepsilon =dK_j^\varepsilon$, as
for example the one 
given in the appendix of \cite{Lechner_Marchetti}, where $J_j^\varepsilon$
is cohomologically equivalent to $J_j$ and regular on $M_j$. Then
$d\left(J_iK_j\right)=
\lim_{\varepsilon\rightarrow 0}d\left(J_i K_j^\varepsilon\right) 
=\lim_{\varepsilon\rightarrow 0}J_i J_j^\varepsilon$, and as shown
in \cite{cinesi}, $J_i J_j^\varepsilon$ is cohomologically equivalent
to $J_{ij}\chi_{ij}$ for every $\varepsilon$. This proves that the l.h.s. of
\eref{eq:fundamental_id} 
is {\it cohomologically} equivalent to  $J_{ij}\chi_{ij}$, and it 
fixes the proportionality coefficient of our {\it local} i.e. point--wise
derivation to unity.

\section{Appendix: Equations of motion for the $U(1)$ theory on the
  $D$--branes} 
The equations of motion obtained by the action \eref{eq:action2}  for the $U(1)$
fields are: 
 
\ba 
	d * \htw_{2} &=& - R_1 , \\ 
	d * \htw_{4} &=& + R_3 - \frac{1}{\gamma} R_1 \Yb_{4,2} -
	\frac{g_6}{2} J_8 \frac{A_{4} - A_{8}}{2\pi} , \\ 
	d * \htw_{6} &=& - R_5 + \frac{1}{\gamma} R_3 \Yb_{6,2} -
	\frac{1}{\gamma^2} R_1 \Yb_{6,4} + 
	\frac{g_4}{3} J_8 \frac{A_{6} - A_{8}}{2\pi} \frac{F_{6} -
	F_{8}}{2\pi}    , \\ 
	d * \htw_{8} &=& + R_7 - \frac{1}{\gamma} R_5 \Yb_{8,2} + 
	\frac{1}{\gamma^2} R_3 \Yb_{8,4} - \frac{1}{\gamma^3} R_1
	\Yb_{8,6}  \nonumber \\   &&  - g_2 \left[ 
	\frac{1}{2} J_4 \frac{A_{8} - A_{4}}{2\pi} - \frac{1}{3} J_6 
	\frac{A_{8} - A_{6}}{2\pi} \frac{F_{8} -F_{6}}{2\pi} 
	\right]  . 
\ea 
On the right hand side the regularized products of currents and
	Chern-kernels are always understood.

\section{Appendix: the Wess--Zumino term}
 
Here we make contact between the Wess-Zumino written as the integral
of an $11$--form \eref{eq:WZ1} and the one written in usual ten
dimensional notation, \eref{eq:L10_C} and \eref{eq:L10_rem}. The procedure
one realizes in practice is the following: first of all construct
\eref{eq:L10_C}, that is completely fixed by equations of motion as
showed in section \eref{sec:action}. Then, the remaining part
\eref{eq:L10_rem} is completely fixed by asking invariance of the
action under $Q$ and $P$--transformations. The actual calculations are
lenghty, though straightforward, and we do not include them here. Once
the ten dimensional Wess-Zumino is fixed, one can take its
differential and get the much simpler form \eref{eq:WZ1}, which
displays all invariances and anomaly cancellations at a first
sight. What we do here instead is to proceed in the opposite direction, that
is to show how to transform \eref{eq:WZ1} into the ten dimensional
Wess-Zumino. 
 
As a first step, consider \eref{eq:WZ1} and extract all the terms
dependent on the potentials. After some algebra and integration by
part one shows that this amounts to 
\be 
	d \left[ -\frac{1}{2} C_4 \left( R_3 H +\Delta_6  \right)
	- \frac{1}{2} C_2 H f_5 - C_2 \Delta_8 - C_0 \Delta_{10}
	\right] = d L_{10}^C . 
\ee 
Now, take the remainder in \eref{eq:WZ1}. This is equal to 
\be 
	\begin{array}{ll} 
	+& \frac{1}{2} f_5 (f_3 H - \Delta_6 ) + f_3 \Delta_8 - f_1
	\Delta_{10} - G P_{11}  \\ = &  
	\frac{1}{2} d (f_3 f_7 - f_1 f_9 ) + \frac{1}{2} \left(
	-\Delta_2 f_9 + \Delta_4 f_7 - \Delta_6 f_5 + \Delta_8 f_3 -
	\Delta_{10}  f_1 \right) - G P_{11}  \\ = &   
	\frac{1}{2} d (f_3 f_7 - f_1 f_9 ) + \frac{1}{2}
	\Sigma_{n} (-1)^{\frac{n}{2}+1}  \Delta_{10-n} f_{n+1} - G P_{11}  \\
	= &  \frac{1}{2} d (f_3 f_7 - f_1 f_9 ) + \frac{1}{2} \left. 
	\Sigma_{n} (-1)^{\frac{n}{2}+1} \Delta_{10-n} f_{n+1}\right|_{B_2=0}
	- G P_{11} , 
	\end{array} \label{eq:boh} 
\ee 
where in the last passage independence from $B_2$ depends crucially on
the alternating sign and can be checked using \eref{eq:delta_Delta}
and an analogous variation for $f$. Given this, one shows with some
algebra that 
\be 
	\left. 
	 \Sigma_{n} (-1)^{\frac{n}{2}+1}  \Delta_{10-n}
	f_{n+1}\right|_{B_2=0} =  2\pi G  
	\sum_{i,j} (-1)^{i/2+1} P_{ij} K_i J_j . \label{eq:boh2} 
\ee 
Putting together eqs.\eref{eq:boh}, \eref{eq:boh2} and the expression
\eref{eq:P11} for $P_{11}$ one gets that the reminder is exactly given
by \eref{eq:L10_rem}. 
 
\bigskip

\paragraph{Acknowledgements.}
The authors thank P.A. Marchetti for useful discussions.
This work is supported in part by the European Community's Human
Potential Programme under contract HPRN-CT-2000-00131 Quantum
Spacetime. M. C. is funded by EPSRC, Fondazione Angelo Della Riccia, Firenze and 
Cambridge European Trust.

\end{document}